\documentclass[prd,twocolumn,aps,floats,showpacs]{revtex4}
\usepackage{graphicx}
\usepackage{bm}
\usepackage{epsfig}
\usepackage{amsmath,amsfonts}
\usepackage{url}

\begin{document}

\title{Jeans analysis in modified gravity}
\author{Mahmood Roshan$^{1}$\footnote{mroshan@um.ac.ir} and
Shahram Abbassi$^{1,2,3}$\footnote{abbassi@ipm.ir}}
\affiliation{$^{1}$Department of Physics, Ferdowsi University of Mashhad, P.O. Box 1436, Mashhad, Iran}
\affiliation{$^{2}$ School of Astronomy, Institute for Research in Fundamental Sciences (IPM), P.O. Box 19395-5531, Tehran, Iran}
\affiliation{$^{3}$Key Laboratory for Research in Galaxies and Cosmology, Shanghai Astronomical Observatory, Chinese Academy of Sciences, 80 Nandan Road, Shanghai 200030, China}
\date{\today}
\begin{abstract}
MOdified Gravity (MOG) is a covariant modification of Einstein's general relativity. This theory is one of the current alternatives to dark matter models. We describe dynamics of
collisionless self-gravitating systems in the context of MOG. By studying the weak field approximation of this theory, we derive the equations governing the dynamics of the
self-gravitating systems. More specifically, we consider the Jeans instability for self-gravitating fluid and stellar systems, and derive new Jeans mass
limit $\tilde{M}_J$ and wave-number $\tilde{k}_J$. Furthermore, considering the gravitational instability in star forming regions, we show that MOG has not a significant difference with general relativity on this astrophysical scale. However, at larger scales such as intergalactic space MOG may lead to different galaxy and structure formation processes.
\end{abstract}


\maketitle

 \section{Introduction}

 It is known that without introducing dark matter and dark
energy, General Relativity (GR) would not be a viable gravitation
theory at the astrophysical and cosmological scales.
In the context of the currently favored cosmological model, the $\Lambda$ cold dark matter
($\Lambda$CDM) model, dark matter is the dominating mass component of the universe.
In the $\Lambda$CDM model, it is believed that 4\% of the mass-energy content of the universe is in the form of ordinary matter (baryonic matter) while 96 \% is in an unknown dark
form (about 22 \% being the exotic cold dark matter and about 74\% being in dark energy). Furthermore, without cold dark matte particles the current structures of
the universe can not be promoted. On the  other hand, the possible
connection with extensions of the Standard Model of particle physics makes
dark matter one of the most complex scandals of cosmology and particle
physics.

In addition, it should be mentioned that despite of several evidences at astronomical and cosmological scales, dark matter particles has not been
detected after much experimental efforts using different techniques and
technologies \cite{berton}. Therefore, nature of dark matter remains unknown. Recently, as one of the most sensitive experiments,
the LUX experimental group, has announced new upper limits on the cross
section of WIMP-nucleon elastic scattering \cite{lux}. This result rules out
earlier suggestions of low mass WIMP detection signals \cite{LUX2}. In fact, this
group has continued the previous negative results for direct detection of dark matter particles.

An alternative framework to dark matter problem is modified gravity. The aim of modified gravity theories is to address the
dark matter problem without invoking new particles but with introducing
some extensions to GR. One promising alternative theory of gravity is MOdified Gravity (MOG) of Moffat
\cite{mofi}. MOG is a Scalar-Tensor-Vector Gravity (STVG) \cite{mofi}. In this theory along with metric tensor there are
three scalar gravitational fields and one dynamical massive vector field.
MOG has recently been successfully applied for the rotation curves of spiral galaxies \cite{bro1,rahvar} and the
observed mass discrepancy in galaxy clusters \cite{bro2}. Also, this theory has
been used to explain the interesting dynamics of the so-called
Bullet cluster 1E0657-558 \cite{mofi3}.

In this paper, we analyze the Jeans instability for self-gravitating fluid and stellar systems in MOG. Our motivation is the significant application of the Jeans analysis in the star formation scenarios and also in the theory of structure formation. In order to do so, we need the weak field limit of the theory.
The weak field approximation of MOG has already been investigated in the literature, for example see \cite{mofi,rahvar}. However, we briefly review the subject and obtain the weak
field limit of MOG. We also compare our results with those previously reported in the literature and discuss the differences. Using the weak field limit, we investigate the dynamics of self-gravitating systems involved in star formation. The main purpose is to compare MOG with GR at astrophysical issues. This comparison combined with the related observational data may provide an evidence for viability of MOG at astrophysical scale.

It is worth to mention that a large number of papers can be found in the literature considering the Jeans analysis and related subjects such as collapsing stars, and black holes in the context of modified gravity theories. For example, the Jeans analysis of the self-gravitating systems in $f(R)$ gravity has been investigated in \citep{capo,capo100,dolgov}. For papers about spherical collapsing stars, and black holes in $f(R)$ gravity see \cite{clifton,dombriz}. Also for papers considering the spherical collapse in the context of modified Newtonian dynamics (MOND) see \cite{haghi}.

The layout of the paper is the following. In section \ref{Newton}, the weak field approximation of MOG is obtained and differences with previous results is discussed.
Also, the coupled Boltzmann and Poisson equations are derived. In fact, we find the correction induced by MOG to the Newtonian gravitational potential. Obviously, this correction would produce
a correction to the collisionless Boltzmann equation. In section \ref{fluid} and \ref{stellar}, by linearizing the equations, we study the Jeans instability for fluid and stellar
systems. The main result of these two sections is the "generalized" Jeans mass limit. In section \ref{MT}, some relevant astrophysical systems are considered. More specifically, searching for the differences of MOG with GR in the star formation, galaxy and structure formation issues, we compare the new Jeans mass with the standard one in the interstellar and intergalactic medium. Finally, in section
\ref{DC}, results are discussed.

\section{Weak field limit of MOG: Poisson and Boltzmann equations}\label{Newton}

As we mentioned in the introduction, MOG is a scalar-tensor-vector gravity theory which postulates, in addition to the metric field, three scalar fields and also a
dynamical massive four-vector gravitational field. The vector field is coupled to matter and causes a nonzero covariant divergence for the ordinary matter energy-momentum
tensor. Consequently there is a violation of Einstein's Equivalence Principle \cite{PRDme}. Let us start with the generic action of MOG
\begin{equation}
S=S_{\phi}+S_{\psi}+S_{\omega}+S_{\chi}+S_{Grav}+S_{M}
\label{mog1}
\end{equation}
where $S_{M}$ is the action for ordinary matter and
\begin{equation}
S_{\phi}=-\int \sqrt{-g}~ d^4 x ~\omega\left[\frac{1}{4}B^{\mu\nu}B_{\mu\nu}-\frac{e^{2\psi}}{2} \phi_{\alpha}\phi^{\alpha}+V_{\phi}\right]
\label{mog3}
\end{equation}

\begin{equation}
S_{\psi}=\int \sqrt{-g}~ d^4 x ~\frac{\chi^2}{2}\left[\frac{1}{2}g^{\mu\nu}\nabla_{\mu}\psi\nabla_{\nu}\psi-V_{\psi}\right]
\label{mog5}
\end{equation}
\begin{equation}
S_{\omega}=\int \sqrt{-g}~ d^4 x ~\frac{\chi^2}{2}\left[\frac{1}{2}g^{\mu\nu}\nabla_{\mu}\omega\nabla_{\nu}\omega-V_{\omega}\right]
\label{mog6}
\end{equation}
\begin{equation}
S_{\chi}=\int \sqrt{-g}~ d^4 x ~\left[\frac{1}{2}g^{\mu\nu}\nabla_{\mu}\chi\nabla_{\nu}\chi-V_{\chi}\right]
\label{mog4}
\end{equation}
\begin{equation}
S_{Grav}=\frac{1}{16\pi}\int \sqrt{-g}~ d^4 x ~\frac{\chi^2}{2} R
\label{mog2}
\end{equation}
the signature is $(-,+,+,+)$, $R$ is the Ricci scalar, $B_{\mu\nu}=\nabla_{\mu}\phi_{\nu}-\nabla_{\nu}\phi_{\mu}$ and the scalar fields $\psi$ and $\chi$ are related to the scalar fields
 $\mu$ and $G$ introduced in \cite{mofi} by
\begin{equation}\label{jmog1}
\frac{\chi^2}{2}=\frac{1}{G},~~~~~~~~~~~\psi=\ln \mu
\end{equation}
also $V_{\chi}$ and $V_{\psi}$ are related to $V_{G}$ and $V_{\mu}$ by
\begin{equation}
    V_{G}=V_{\chi} G^3~~~~~~~~~~~V_{\mu}=V_{\psi}\mu^2
\end{equation}
The total energy-momentum tensor is given by
\begin{equation}
   T^{total}_{\mu\nu}=T_{\mu\nu}+T^{\phi}_{\mu\nu}+T^{\chi}_{\mu\nu}+T^{\psi}_{\mu\nu}+T^{\omega}_{\mu\nu}
\end{equation}
One can then verify that the covariant divergence of the matter energy-momentum tensor is given by the following expression (for details see \cite{PRDme})
\begin{equation}\label{mog23}
       \nabla^{ \mu}T_{\mu\nu} =B_{\alpha\nu}J^{\alpha}+\nabla^{\mu}\left(g_{\mu\alpha}\omega\phi_{\nu}\frac{\partial V_{\phi}}{\partial \phi_{\alpha}}-2\omega
       \frac{\partial V_{\phi}}{\partial g^{\mu\nu}}\right)
\end{equation}
where $J^{\alpha}$ is a "fifth force" matter current defined as
 \begin{equation}\label{current1}
J^{\alpha}=-\frac{1}{\sqrt{-g}}\frac{ \delta S_M}{\delta \phi_{\alpha}}
 \end{equation}

Variation of the action \eqref{mog1} with respect to $g^{\mu\nu}$, $\chi$, $\phi_{\alpha}$, $\psi$ and $\omega$, and taking into account equation \eqref{jmog1},
yields to the following field equations respectively (for more details see \cite{cqg2009})
 \begin{widetext}
 \begin{equation}
 \begin{split}
R_{\mu\nu}-\frac{1}{2}g_{\mu\nu}R&+g_{\mu\nu}\left(\frac{2}{G^2}\nabla_{\alpha}G\nabla^{\alpha}G-\frac{\square G}{G}+G\omega\mu^2 \phi_{\alpha}\phi^{\alpha}-4\pi\frac{\nabla_{\alpha}\mu\nabla^{\alpha}\mu}{\mu^2}+4\pi\nabla_{\alpha}\omega\nabla^{\alpha}\omega \right)-\frac{2\nabla_{\mu}G\nabla_{\nu}G}{G^2}\\& + \frac{\nabla_{\mu}\nabla_{\nu}G}{G}-8\pi\left(\frac{1}{4\pi}G\omega\mu^2\phi_{\alpha}\phi^{\alpha}-\frac{\nabla_{\mu}G\nabla_{\nu}G}{G^2}-
\frac{\nabla_{\mu}\mu\nabla_{\nu}\mu}{\mu^2}+\nabla_{\mu}\omega\nabla_{\nu}\omega\right)\\&
+\frac{1}{4\pi}G\omega\left(B^{\alpha}_{~\mu}B_{\nu\alpha}+\frac{1}{4}B^{\alpha\beta}B_{\alpha\beta}g_{\mu\nu}\right)=-8\pi G T_{\mu\nu}
\end{split}
\label{mog11}
\end{equation}
 \begin{equation}
    \square G+\frac{G}{2} \left(\frac{\nabla^{\alpha}\mu\nabla_{\alpha}\mu}{\mu^2}-\nabla_{\alpha}\omega\nabla^{\alpha}\omega\right)
    -\frac{3}{2G}\nabla^{\alpha}G\nabla_{\alpha}G+\frac{G}{16\pi}R=0
 \label{mog13}
  \end{equation}
 \end{widetext}

 \begin{equation}\label{mog12}
    \omega \nabla_{\mu}B^{\mu\nu}+\nabla_{\mu}\omega B^{\mu\nu}+\omega \mu^2\phi^{\nu}=4\pi J^{\alpha}
 \end{equation}

 \begin{equation}\label{mog14}
    \square \mu-\frac{\nabla_{\alpha}\mu\nabla^{\alpha}\mu}{\mu}-\frac{\nabla_{\alpha }\mu\nabla^{\alpha}G}{G}+\frac{1}{4\pi}G\omega\mu^3\phi^{\alpha}\phi_{\alpha}=0
 \end{equation}
 \begin{equation}\label{mog15}
     \square \omega-\frac{\nabla^{\alpha}G\nabla_{\alpha}\omega}{G}-\frac{1}{8\pi}G\mu^2\phi^{\alpha}\phi_{\alpha}+\frac{G}{16\pi}B^{\mu\nu}B_{\mu\nu}=0
 \end{equation}
 where for the sake of simplicity, we have assumed that self-interaction potentials $V_{\phi}$, $V_{\psi}$, $V_{\chi}$ and $V_{\omega}$ are zero. Also, we assume the adiabatic approximation in which the evolution of the universe is very slow in comparison with local dynamics. This assumption allows us to choose the Minkowski space-time instead of the Friedmann-Robertson-Walker (FRW) space-time as the background metric (see \cite{clifton} and references cited therein). Thus, in order to write the field equations in the first perturbation order, we decompose metric into the Minkowski metric $\eta_{\mu\nu}$ plus a small perturbation $h_{\mu\nu}$ ($|h_{\mu\nu}|\ll 1$) 
 \begin{equation}\nonumber
 g_{\mu\nu}=\eta_{\mu\nu}+h_{\mu\nu}=\left(
              \begin{array}{cc}
                1+2 \Phi(\mathbf{x},t) & 0 \\
                0& (-1+2\Psi(\mathbf{x},t))\delta_{ij}\\
              \end{array}
            \right)
 \end{equation}
We will confine ourselves to coordinates in which $\eta_{\mu\nu}$ takes its canonical form $\eta_{\mu\nu}=\text{diga}(+1,-1,-1,-1)$. Also we use the \textit{transverse gauge}, i.e.
\begin{equation}
\partial_i s^{ij}=0
\end{equation}
where the traceless tensor $s_{ij}$ is known as the \textit{strain} and is given by
\begin{equation}
s_{ij}=\frac{1}{2}\left(h_{ij}-\frac{1}{3}\delta^{kl}h_{kl}\delta_{ij}\right)
\end{equation} 
furthermore perturbations of the other fields are
  \begin{equation}
 \begin{split}
 & \phi^{\mu}=\phi_{0}^{\mu}+\delta\phi^{\mu}(\mathbf{x},t)\\&
            G=G_{0}+\delta\ G (\mathbf{x},t)\\&
            \omega=\omega_0+\delta\omega(\mathbf{x},t)\\&
            \mu=\mu_{0}+\delta\mu(\mathbf{x},t)\\&
            J^{\mu}=J_{0}^{\mu}+\delta J^{\mu}(\mathbf{x},t)
 \end{split}
 \label{pert}
 \end{equation}
where $\phi^{\mu}_{0}$, $G_0$, $\omega_{0}$, $J_{0}^{\mu}$ and $\mu_0$ are the background value of the fields. Also $\delta\phi^{\mu}$, $\delta G$, $\delta \omega$, $\delta J^{\mu}$ and $\delta\mu$ are the perturbations around the Minkowski background.
 Using the field equations it is easy to verify that the background values $\phi^{\mu}_{0}$ and $J_{0}^{\mu}$ are zero.

Preserving only the first-order perturbations, the metric field equation \eqref{mog11} takes the following form
\begin{equation}
\delta G_{\mu\nu}+\eta_{\mu\nu}\frac{\nabla^{2}\delta G}{G_0}+\frac{\partial_{\mu}\partial_{\nu}\delta G}{G_0}=-8\pi G_0 T_{\mu\nu}
\label{jmog2}
\end{equation}
where $G_{\mu\nu}$ is the Einstein tensor. On the other hand, one can verify that
\begin{equation}
\delta G_{\mu\nu}\sim \left(
                                                         \begin{array}{cc}
                                                           -2\nabla^2\Psi & 0 \\
                                                           0 & \left( \delta_{ij} \nabla^2-\partial_i\partial_j\right)(\Psi-\Phi) \\
                                                         \end{array}
                                                       \right)
\end{equation}
Taking into account that for a perfect fluid $T_{ij}$ and $T_{0j}$ are zero in the Newtonian limit, and considering all of the components of Eq.
\eqref{jmog2}, we find the following equations for $\Psi$ and $\Phi$
\begin{equation}
2\nabla^{2}\Psi-\frac{\nabla^{2}\delta G}{G_0}=8\pi G_0 T_{00}
\label{jmog3}
\end{equation}
\begin{equation}
\nabla^2\Phi-\nabla^2\Psi+\frac{\nabla^{2}\delta G}{G_0}=0
\label{jmog4}
\end{equation}
on the other hand the trace of Eq. \eqref{jmog2} yields
\begin{equation}
\nabla^{2}\delta G=\frac{8\pi G_0^2}{16\pi-3}T
\label{jmog5}
\end{equation}
combining equations \eqref{jmog3}-\eqref{jmog5} and taking into account that in the Newtonian limit $T_{00}=T\sim \rho$, we find
\begin{equation}
\nabla^2 \Psi=4\pi G_0 \left(\frac{16\pi-2}{16\pi-3}\right)\rho
\label{jmog6}
\end{equation}
\begin{equation}
\nabla^2 \Phi=4\pi G_0 \left(\frac{16\pi}{16\pi-3}\right)\rho
\label{jmog60}
\end{equation}
Equations \eqref{jmog5} and \eqref{jmog6} are different from corresponding equations (19) and (21) in \cite{rahvar}. Strictly speaking, equations \eqref{jmog5} and \eqref{jmog6} are similar to the Poisson's equation. However this similarity does not mean that MOG behaves like GR in the weak field limit. In the next section, we show that the modified Poisson equation in MOG is totally different from the standard Poisson equation.

 One can simply find the solutions of Eqs. \eqref{jmog6} and \eqref{jmog60}
\begin{equation}
\begin{split}
& \Psi=-\left(\frac{16\pi -2}{16\pi-3}\right)G_0 \int \frac{\rho(\mathbf{x'})}{|\mathbf{x}-\mathbf{x'}|}d^3x'\\& \Phi=-\left(\frac{16\pi }{16\pi-3}\right)G_0 \int \frac{\rho(\mathbf{x'})}{|\mathbf{x}-\mathbf{x'}|}d^3x'\\
\end{split}
\end{equation}
Thus there is a relation between metric components $\Psi$ and $\Phi$
\begin{equation}
\Psi=\frac{16\pi-2}{16\pi}\Phi
\label{gamma}
\end{equation}
In the context of metric theories of gravity, where the validity of Einstein's Equivalence Principle is postulated \cite{willbook}, the post-Newtonian parameter $\gamma_{PPN}$ is defined as
$\gamma_{PPN}=\frac{\Psi}{\Phi}$. It is worth to mention that in metric theories of gravity,
in order to pass the classical tests of GR $\gamma_{PPN}$ should satisfy the limit
$\gamma_{PPN}=1 + (2.1 \pm2.3) \times 10^{-5}$ \cite{will}. We know that MOG is not a metric theory because there is a
coupling between $\phi^{\mu}$ and matter. As we mentioned before, this coupling causes a violation of the Einstein's Equivalence Principle. Thus, we expect
that the magnitude of $\gamma_{MOG}$ in MOG would not necessarily satisfy the above-mentioned limit. However, it is worth to mention the magnitude of $\gamma_{MOG}$ in MOG. Using Eq. \eqref{gamma} one
can straightforwardly write $\gamma_{MOG}=\frac{\Psi}{\Phi}=1-\frac{2}{16\pi}=0.9602$.

Keeping only first order terms in the other field equations, we obtain
\begin{equation}
\nabla^{2}\delta\phi^{\mu}-\mu_0 \delta\phi^{\mu}=-\frac{4\pi}{\omega_0}\delta J^{\mu}
\label{jmog10}
\end{equation}
\begin{equation}
\nabla^2\delta G=\frac{8\pi G_0^2}{16\pi-3}\rho
\label{jmog7}
\end{equation}
\begin{equation}
\nabla^2 \delta\mu=0
\end{equation}
\begin{equation}
\nabla^2 \delta\omega=0
\label{jmog8}
\end{equation}
Equations \eqref{jmog6}-\eqref{jmog8} are the main equations required for considering a self-gravitating system in the weak filed approximation in the context of MOG. In the following sections
with the aid of these linearized equations, we derive the generalized version of the Poisson and Boltzmann equations.
\subsection{The Poisson equation in MOG}

In order to find the MOG Poisson equation, let us start with the equation of motion of a test particle
\begin{equation}
\frac{d^2x^i}{d\tau^2}+\Gamma^{i}_{\alpha\beta}\frac{dx^{\alpha}}{d\tau}\frac{dx^{\beta}}{d\tau}=\kappa B^{i}_{~\alpha}\frac{dx^{\alpha}}{d\tau}
\label{jmog9}
\end{equation}
where $\kappa$ is a coupling constant and its magnitude can be specified with experimental observations, and $\tau$ is the proper time along the world line
of the test particle. It is worth to mention that in GR it is not needed to postulate the equation of
motion of a test particle as an independent equations of the field equations. In other words, the geodesics equation can be directly derived from the field equations \cite{papa}. This is also the case in MOG, and 
the equation of motion \eqref{jmog9} can be derived from the field equations of MOG \cite{PRDme}. It is straightforward to show that the governing action for the test particle motion is given by
\begin{equation}
S_{TP}=-\int m\left(1+\kappa\phi_{\mu}\frac{dx^{\mu}}{d\tau}\right)d\tau
\end{equation}
where $m$ is the test particle's rest mass. In the literature on MOG, it is usually assumed that the scalar field $\omega$ appears directly in the test particle action
alongside the massive vector field $\phi^{\mu}$ (for example see \cite{mofi} and \cite{rahvar}). Presence of the vector field $\phi^{\mu}$ in the action is expected since
there is a coupling between it and ordinary matter. On the other hand, there is also a coupling between $\phi^{\mu}$
 and $\omega$, see equation \eqref{mog3}. Thus, one may expect that the scalar field $\omega$ should appear also in the equation of motion. However, this is not
 the case and using the generalized Mathisson-Papapetrou equations one can show that $\omega$ does not appear in the test particle action \cite{PRDme}.

Taking into account the above considerations, the action for pressureless dust can be written as
\begin{equation}
S_{M}=-\int \rho\left(1+\kappa\phi_{\mu}\frac{dx^{\mu}}{d\tau}\right)\sqrt{-g}d^4x
\end{equation}
where $\rho$ is density of matter. Thus, using the above action and Eq. \eqref{current1}, one can easily verify that the time component of the fifth-force
 matter current is $J^{0}=\kappa \rho$.

Now, it is straightforward to write Eq. \eqref{jmog9} to first order in perturbations
\begin{equation}
\frac{d^2\mathbf{r}}{dt^2}=-\nabla \Phi_{eff}
\label{new0}
\end{equation}
where the effective potential $\Phi_{eff}$ is defined as
\begin{equation}
\Phi_{eff}=\Phi+\kappa \delta\phi^{0}
\label{jmog12}
\end{equation}
On the other hand, using Eq. \eqref{jmog10} we arrive at
\begin{equation}
\delta\phi^{0}=\frac{1}{\omega_0}\int \frac{e^{-\mu_0|\mathbf{x}-\mathbf{x'}|}}{|\mathbf{x}-\mathbf{x'}|}\delta J^{0}(\mathbf{x'})d^3x'
\label{jmog11}
\end{equation}
by combining Eqs. \eqref{jmog60}, \eqref{jmog12} and \eqref{jmog11}, and setting $\delta J^{0}=\kappa\rho$, we arrive at the modified version of the Poisson equation
\begin{equation}
\nabla^2{\Phi_{eff}}=4\pi (G_0' -\frac{\kappa^2}{\omega_0})\rho+\frac{\kappa^2\mu_0^2}{\omega_0} \int \frac{e^{-\mu_0|\mathbf{x}-\mathbf{x'}|}}{|\mathbf{x}-\mathbf{x'}|}\rho(\mathbf{x'})d^3x'
\end{equation}
where
\begin{equation}
G_0'=\frac{16\pi G_0}{16\pi-3}
\end{equation}
the solution for the effective potential is
\begin{equation}
\Phi_{eff}=-G_0'\int \frac{\rho(\mathbf{x'})}{|\mathbf{x}-\mathbf{x'}|}d^3x'+\frac{\kappa^2}{\omega_0} \int \frac{e^{-\mu_0|\mathbf{x}-\mathbf{x'}|}}{|\mathbf{x}-\mathbf{x'}|}\rho(\mathbf{x'})d^3x'
\end{equation}
It is obvious that there are four free parameters in $\Phi_{eff}$. However, by comparing this effective potential with the Newtonian gravitational potential,
one can find a relation between $G_0$, $\kappa$, $\omega_0$ and $G_N$ ($G_N$ is the Newton's gravitational constant).
To do so, let us consider the effective gravitational potential produced by a point mass  $M$. In this case, the density is $\rho(x)=M\delta^3(\mathbf{x})$.
Thus, the effective potential is
\begin{equation}
\Phi_{eff}=-\frac{M G_0'}{r}+\frac{\kappa^2}{\omega_0}\frac{M e^{-\mu_0 r}}{r}
\label{new1}
\end{equation}
at small distances, we expect that $\mu_0 r \ll 1$ and also $\Phi_{eff}\simeq \Phi_N=- G_N M/r$. Thus, by expanding the exponential term, one obtains the following algebraic
 relation between the parameters
\begin{equation}
G_0'-\frac{\kappa^2}{\omega_0}=G_N
\label{new2}
\end{equation}
Now, it seems that there are three independent free parameters, for example $\mu_0$, $\kappa$ and $\omega_0$. However, $\kappa$ and $\omega_0$ do not appear separately
in the Poisson equation and effective potential. It is clear that they appear by the combination $\kappa^2/\omega_0$. In practice, one may conclude that there are
only two independent parameters $\mu_0$ and $\lambda=\kappa^2/\omega_0$ which should be determined by the relevant observations.
  Following \cite{mofi}, instead of using $\lambda$, we shall use parameter $\alpha$ defined as $\alpha =\kappa^2/\omega_0 G_N$. Finally, we rewrite the modified Poisson
  equation and the effective potential as follows
\begin{equation}
\nabla^2\Phi_{eff}=4\pi G_N \rho +\alpha\mu_0^2 G_N \int \frac{e^{-\mu_0|\mathbf{x}-\mathbf{x'}|}}{|\mathbf{x}-\mathbf{x'}|}\rho(\mathbf{x'})d^3x'
\label{poissone}
\end{equation}
\begin{equation}
\Phi_{eff}=-G_N\int \frac{\rho(\mathbf{x'})}{|\mathbf{x}-\mathbf{x'}|}\left(1+\alpha-\alpha e^{-\mu_0|\mathbf{x}-\mathbf{x'}|}\right)d^3x'
\end{equation}
For $\alpha=0$ or $\mu_0=0$ the standard Poisson equation is recovered. It should be stressed that although some of our equations are different from those of \cite{rahvar}, but because of
the freedom to define parameters $G_0'$ and $\alpha$, our final result for Poisson equation coincides with that of \cite{rahvar}.

It is also worth to mention that, we assume $\omega_0>0$ and consequently $\alpha>0$. This assumption is necessary in the sense that in order to address the dark matter problem, the gravitational force predicted by MOG in the weak field limit should be stronger than that of Newtonian limit of GR. See equation \eqref{new3} and the subsequent paragraph.
\subsection{The collisionless Boltzmann equation in MOG}
In this section, we find the collisionless Boltzamnn equation. This equation is required to study the dynamics of a self-gravitating stellar system. Let $f(x^{\mu}, u^{\mu})$
be the particle's phase space distribution function. In Newtonian dynamics the Boltzmann equation can be written as $\hat{L}[f]=0$, where $\hat{L}$ is the Liouville operator.
Using the test particle equation of motion \eqref{jmog9}, the covariant, relativistic generalization of the Boltzmann equation in the context of MOG can be written as
\begin{equation}
\hat{L}[f]=u^{\mu}\frac{\partial f}{\partial x^{\mu}}+\left(\kappa B^{\mu}_{~\alpha}u^{\alpha}-\Gamma^{\mu}_{\alpha\beta}u^{\alpha}u^{\beta}\right)\frac{\partial f}{\partial u^{\mu}}=0
\end{equation}
One can easily linearize this equation using the perturbations introduced in Eq. \eqref{pert}. Taking into account that in the first order Newtonian limit
$\Gamma^i_{00}\simeq \frac{\partial\Phi}{\partial x^i}$, $u^i\simeq v^i=\frac{dx^i}{dt}$ and $B^{i}_{\alpha}u^{\alpha}\simeq\frac{\partial \phi_0}{\partial x^i}$, we obtain
\begin{equation}
\frac{\partial f}{\partial t}+ \mathbf{v}\cdot\nabla f-\nabla \Phi_{eff}\cdot\frac{\partial f}{\partial \mathbf{v}}=0
\label{bolt}
\end{equation}
By comparing this equation with the standard Boltzmann equation, it is obvious that the only difference is the replacement of $\Phi_{eff}$ with $\Phi_N$.
\section{The Jeans instability for a self-gravitating fluid in  MOG}\label{fluid}
We now consider the Jeans stability for a self-gravitating perfect fluid with density $\rho(\mathbf{x},t)$, pressure $p(\mathbf{x},t)$ and velocity $\mathbf{v}(\mathbf{x},t)$. Writing
the divergence of the energy-momentum tensor, Eq. \eqref{mog23}, in the Newtonian limit, one can easily show that the continuity and Euler's equations are
\begin{equation}
\frac{\partial \rho}{\partial t}+\nabla\cdot (\rho \mathbf{v})=0
\label{jmog13}
\end{equation}
\begin{equation}
\frac{\partial \mathbf{v}}{\partial t}+(\mathbf{v}\cdot \nabla)\mathbf{v}=-\frac{1}{\rho}\nabla p-\nabla\Phi_{eff}
\label{jmog14}
\end{equation}
Eqs. \eqref{jmog13},\eqref{jmog14} and \eqref{poissone} together with the equation of state relating $p$ and $\rho$, constitute a set of equations governing the dynamics of a
 self-gravitating fluid. Let us assume that the equilibrium state of a barotropic fluid (i.e. a fluid for which $p=p(\rho)$) is described by $\rho_0({\mathbf{x}})$, $p_0({\mathbf{x}})$,
 $\mathbf{v_0}({\mathbf{x})}$ and $\Phi_{0}({\mathbf{x}})$. Hereafter, for the sake of brevity, we use $\Phi$ instead of $\Phi_{eff}$ and $G$ instead of $G_N$.

 Let us now consider the following small perturbations to the equilibrium state
\begin{equation}
\begin{split}
&\rho(\mathbf{x},t)=\rho_0({\mathbf{x}})+\epsilon \rho_1(\mathbf{x},t)\\&
p(\mathbf{x},t)=p_0({\mathbf{x}})+\epsilon p_1(\mathbf{x},t)\\&
\mathbf{v}(\mathbf{x},t)=\mathbf{v}_0({\mathbf{x}})+\epsilon \mathbf{v}_1(\mathbf{x},t)\\&
\Phi(\mathbf{x},t)=\Phi_0({\mathbf{x}})+\epsilon \Phi_1(\mathbf{x},t)
\end{split}
\label{jmog16}
\end{equation}
where $\epsilon\ll1$. Substituting perturbations \eqref{jmog16} into Eqs. \eqref{jmog13},\eqref{jmog14} and \eqref{poissone}, and preserving only the first order terms in $\epsilon$, we arrive at
\begin{equation}
\frac{\partial\rho_1}{\partial t}+\nabla\cdot(\rho_0 \mathbf{v}_1)+\nabla\cdot(\rho_1 \mathbf{v}_0)=0
\label{jmog15}
\end{equation}
\begin{equation}
\frac{\partial \mathbf{v}_1}{\partial t}+(\mathbf{v}_0\cdot\nabla)\mathbf{v}_1+(\mathbf{v}_1\cdot\nabla)\mathbf{v}_0=-\nabla h_1-\nabla\Phi_1
\label{jmog17}
\end{equation}
\begin{equation}
\nabla^2\Phi=4\pi G\rho_1+\alpha G \mu_0^2\int \frac{\rho_1(\mathbf{x'})e^{-\mu_0|\mathbf{x}-\mathbf{x'}|}}{|\mathbf{x}-\mathbf{x'}|}d^3x'
\label{jmog18}
\end{equation}
where
\begin{equation}
h_1=c_s^2\frac{\rho_1}{\rho_0}~~~~~~~~~~~~c_s^2(\mathbf{x})=\left(\frac{dp}{d\rho}\right)_{\rho_0}
\end{equation}
With the aid of the so-called \emph{Jeans swindle} (for details see \cite{binney}), we set $\rho_0= constant$ and $\mathbf{v}_0=0$ for an infinite homogeneous fluid.
Combining the time-derivative of Eq. \eqref{jmog15} and the divergence of Eq. \eqref{jmog17}, we get
\begin{equation}
\begin{split}
\frac{\partial^2\rho_1}{\partial t^2}-&c_s^2\nabla^2\rho_1-4\pi G \rho_0 \rho_1\\&-\alpha G \mu_0^2\rho_0 \int \frac{\rho_1(\mathbf{x'})e^{-\mu_0|\mathbf{x}-\mathbf{x'}|}}{|\mathbf{x}-\mathbf{x'}|}d^3x'=0
\end{split}
\end{equation}
In Fourier components, $\rho_1\sim e^{i(\mathbf{k}\cdot\mathbf{x}-\omega t)}$, this equation becomes
\begin{equation}
\begin{split}
\omega^2-c_s^2 k^2+&4\pi G \rho_0\\&+\alpha G \mu_0^2 \rho_0 \int \frac{e^{-\mu_0|\mathbf{x}-\mathbf{x'}|+i \mathbf{k}\cdot(\mathbf{x}-\mathbf{x'})}}{|\mathbf{x}-\mathbf{x'}|}d^3x'=0
\end{split}
\label{jmog19}
\end{equation}
Note that $\omega$ is frequency and should not be confused with the gravitational scalar field $\omega(\mathbf{x})$. Fortunately, the integral appeared in Eq. \eqref{jmog19}
converges and can be readily solved. To do so, in the spherical coordinate system, we impose $\mathbf{k}=k \hat{z}$, where $\hat{z}$ is the unit vector in the direction of the $z$ axis.
Then integrating over all directions $\mathbf{x}$, we obtain the following dispersion relation
\begin{equation}
\omega^2=c_s^2 k^2-4\pi G \rho_0\left(1+\frac{\alpha \mu_0^2}{k^2+\mu_0^2}\right)
\label{jmog20}
\end{equation}
If we set $\alpha=0$ or $\mu_0=0$, then the above equation coincides with the corresponding dispersion relation in Newtonian limit of GR. In order to derive a stability criterion for
 the system, let us define the dimensionless parameter
 \begin{equation}
  \beta=\frac{\mu_0^2}{k_J^2}
 \label{beta}
 \end{equation}
 where $k_J^2=4\pi G \rho_0/c_s^2$ is the Jeans wavenumber. Since the system is unstable
 if $\omega^2<0$ and stable if $\omega^2>0$, by setting $\omega=0$ in Eq. \eqref{jmog20}, the limit for stability is obtained:
\begin{equation}
\tilde{k}^2_J=\frac{k^2_J}{2}\left(1-\beta+\sqrt{(1+\beta)^2+4\alpha\beta}\right)
\label{jmog211}
\end{equation}
Hence the system is stable if
\begin{equation}
k^2>\tilde{k}^2_J
\end{equation}
Equivalently, perturbations with wavelength $\lambda=2\pi/k$ are stable if $\lambda^2<\tilde{\lambda}^2_J$. Since $\alpha$ and $\beta$ are positive parameters, it is clear form Eq.
\eqref{jmog211} that $\tilde{k}^2_J>k^2_J$. In other words, the generalized Jeans wavelength in MOG is smaller than the Jeans wavelength in Newtonian theory. This means that there
are some unstable modes which are not present in the standard Jeans analysis. In order to show this point more clearly, let us derive a relation
 for the Jeans mass in MOG. The new Jeans mass $\tilde{M}_J$ is defined as the mass inside a sphere with diameter $\tilde{\lambda}_J$:
\begin{equation}
\tilde{M}_J=\left(\frac{2}{1-\beta+\sqrt{(1+\beta)^2+4\alpha\beta}}\right)^{3/2}M_J
\label{jmass}
\end{equation}
where $M_J=\pi\rho_0\lambda_J^3/6$ is the standard Jeans mass. It is clear that $\tilde{M}_J$ is smaller than $M_J$. If the mass of the perturbation exceeds the new Jeans mass
$\tilde{M}_J$, it is subject to collapse under its own gravity. This means that some perturbations, which are light enough to be gravitationally stable in Newtonian limit of GR, are
unstable in the context of MOG.

This point can also be expressed in terms of the size of the perturbations. In fact, perturbations with a length scale $D$ longer than $\tilde{\lambda}_J$ are unstable.
In other words, perturbations with $\tilde{\lambda}_J<D<\lambda_J$, which are gravitationally stable in Newtonian limit of GR, are unstable in MOG. Of course, this difference
between Newtonian limit of GR and MOG is directly related to the existence of a correction term in the force law derived form MOG. To see this, using Eqs. \eqref{new0}, \eqref{new1}
and \eqref{new2}, we write the gravitational force exerted on a unit mass by a point mass $M$ as
\begin{equation}
\mathbf{F}=\mathbf{F}_N+\mathbf{F}_N \alpha\left(1-e^{-\mu r}(1+\mu r)\right)
\label{new3}
\end{equation}
where $\mathbf{F}_N=-M G \mathbf{r}/r^3$ is the Newtonian gravitational force. The second term on the right hand side of Eq. \eqref{new3} denotes a force which is attractive at all radii $r$ provided that $\alpha>0$. However, at
small radii this term is small compared to the Newtonian term and Newton's law of gravity is recovered. Therefore this correction makes the gravitational force stronger and consequently the Jeans mass smaller.

 It is worth to mention that, in general, a gravitational force stronger than Newtonian force may provide an explanation for the observed rotation curves, $v(r)$, of spiral galaxies. In fact, since  $v(r)\sim \sqrt{r F}$, stronger force $F$ yields to less rapidly declining rotation profiles. In particular, this is the case in MOG \cite{bro1}. Furthermore in MOND, gravity is modified so as to become stronger at small acceleration scales \cite{milgrom}.

At this point it may be useful to compare MOG with metric $f(R)$ gravity. $f(R)$ gravity is one of the simple modifications of Einstein's general theory of relativity. To construct such a
theory, it is enough to replace the Ricci scalar R with an arbitrary function of it in the Einstein-Hilbert action. This theory provides a possible explanation for the cosmic speed up \cite{capo3}. The Jeans analysis of self-gravitating systems in $f(R)$ gravity has been studied in \cite{capo}. The new Jeans mass for the model of $f(R)$ studied in \cite{capo} is proportional to the standard Newtonian value as $\tilde{M}_J\simeq0.704 M_J$. Therefore, similar to MOG, $\tilde{M}_J$ is smaller than $M_J$. However, in this respect, there is a main difference between these theories. As it is clear form Eq. \eqref{jmass}, $\beta$ depends on the physical properties of the given system. Thus, the coefficient of $M_J$ in the relation of $\tilde{M}_J$ is not the same for all self gravitating systems and varies from system to system. However, this coefficient in the model of $f(R)$ under study in \cite{capo} is the same for all systems and equal to $0.704$.

\section{The Jeans instability for a self-gravitating stellar system in  MOG}
\label{stellar}
We now come to the case of a self gravitating stellar system. In this case, the main equations governing the system are the Poisson equation \eqref{poissone} and the collisionless Boltzmann equation \eqref{bolt}. Note that the integral of distribution function $f(\mathbf{x},\mathbf{v},t)$ over all velocities yields the density $\rho(\mathbf{x},t)$. Let us describe the equilibrium state of an infinite, homogeneous and time-independent stellar system with distribution function $f_0(\mathbf{v})$ and effective gravitational potential $\Phi_0(\mathbf{x})$. Now consider a small perturbation to the equilibrium state
\begin{equation}
\begin{split}
&f(\mathbf{x},\mathbf{v},t)=f_0(\mathbf{v})+\epsilon f_1(\mathbf{x},\mathbf{v},t)\\&
\Phi(\mathbf{x},t)=\Phi_0(\mathbf{x})+\epsilon\Phi_1(\mathbf{x},t)
\end{split}
\label{new4}
\end{equation}
where again $\epsilon\ll 1$. By substituting perturbations \eqref{new4} into Eqs. \eqref{poissone} and \eqref{bolt}, and keeping only the first order terms, we arrive at the following linearized equations respectively
\begin{equation}
\begin{split}
\nabla^2 \Phi_1=&4\pi G \int f_1 d^3 v \\&+\alpha G\mu_0^2 \int \int \frac{e^{-\mu_0|\mathbf{x}-\mathbf{x'}|}}{|\mathbf{x}-\mathbf{x'}|}f_1(\mathbf{x'}, \mathbf{v},t)d^3v d^3x'
\end{split}
\label{jmog21}
\end{equation}

\begin{equation}
\frac{\partial f_1}{\partial t}+\mathbf{v}\cdot \nabla f_1-\nabla \Phi_1\cdot \frac{\partial f_0}{\partial \mathbf{v}}=0
\label{jmog22}
\end{equation}
note that, again, we have used the so-called \emph{Jeans swindle} to set $\Phi_{0}=0$. Now, in order to find a dispersion relation, we insert the Fourier components $f_1=f_a(\mathbf{v})e^{i (\mathbf{k}\cdot\mathbf{x}-\omega t)}$ and $\Phi_1=\Phi_a e^{i (\mathbf{k}\cdot\mathbf{x}-\omega t)}$ into Eqs. \eqref{jmog21} and \eqref{jmog22}. The result is
\begin{equation}
-k^2 \Phi_a=4\pi G\left(1+\frac{\alpha\mu_0^2}{k^2+\mu_0^2}\right)\int f_a(\mathbf{v})d^3 v
\label{jmog23}
\end{equation}
\begin{equation}
(\mathbf{k}\cdot \mathbf{v}-\omega)f_a(\mathbf{v})-\Phi_a \mathbf{k}\cdot \frac{\partial f_0}{\partial \mathbf{v}}=0
\label{jmog24}
\end{equation}
By substituting Eq. \eqref{jmog23} into \eqref{jmog24} and integrating over all velocity space, we obtain the following dispersion relation
\begin{equation}
1+\frac{4\pi G}{k^2}\left(1+\frac{\alpha\mu_0^2}{k^2+\mu_0^2}\right)\int \frac{\mathbf{k}\cdot\frac{\partial f_0}{\partial \mathbf{v}}}{\mathbf{k}\cdot \mathbf{v}-\omega}d^3 v=0
\label{dismog}
\end{equation}
We assume that the equilibrium state of a stellar system is described by the Maxwellian distribution function $f_0(\mathbf{v})$
\begin{equation}
f_0(\mathbf{v})=\frac{\rho_0}{(2\pi \sigma^2)^{3/2}}e^{-v^2/2\sigma^2}
\label{maxwell}
\end{equation}
where $\sigma$ is the velocity dispersion of the particles (or stars). Substituting Eq. \eqref{maxwell} into \eqref{dismog} and imposing $\mathbf{k}=k \hat{z}$, we obtain
\begin{equation}
1-\frac{2\sqrt{2\pi} G}{k\sigma^3}\left(1+\frac{\alpha\mu_0^2}{k^2+\mu_0^2}\right)\rho_0\int_{-\infty}^{+\infty}\frac{v_z e^{-v_z^2/2\sigma^2}}{k v_z-\omega}dv_z=0
\label{jmog25}
\end{equation}
This is the required dispersion relation for a stellar system. By analogy with the fluid system, one may expect that $\omega=0$ is the boundary between stable and unstable modes. Thus, setting $\omega=0$ in Eq. \eqref{jmog25}, we obtain
\begin{equation}
\sigma^2 \tilde{k}_J^2-4\pi G \rho_0\left(1+\frac{\alpha \mu_0^2}{\tilde{k}_J^2+\mu_0^2}\right)=0
\end{equation}
where $\tilde{k}_J=k(\omega=0)$. Thus, we expect that the limit for stability is $k^2>\tilde{k}_J^2$, where
\begin{equation}
\tilde{k}^2_J=\frac{k^2_J}{2}\left(1-\beta+\sqrt{(1+\beta)^2+4\alpha\beta}\right)
\label{jmog26}
\end{equation}
where $k_J^2=\frac{4\pi G \rho_0}{\sigma^2}$ is the standard Jeans wavenumber for a stellar system. Once again, if we set $\alpha=0$, then the Newtonian stability limit is recovered. Mathematically, the new Jeans wavenumber $\tilde{k}_J$ for a stellar system Eq. \eqref{jmog26} is the same as that derived for fluids, see Eq. \eqref{jmog21}. However, one should note that the velocity dispersion $\sigma$ is substituted for the sound speed $c_s$. Thus the new Jeans mass limit for a stellar system is also given by Eq. \eqref{jmass}.

However, let us check that the expectation that all perturbations with wavenumbers $k<\tilde{k}_J$ are unstable, is true in the stellar system. In order to evaluate the integral in the Eq. \eqref{jmog25}, one should study the singularity at $\omega=k v_z$. Obviously, the analysis of the singularity is completely the same as the standard case in Newtonian limit of GR (for details, see Section 5.1 and Appendix 5.A in \cite{binney}). More specifically, as in the standard case, there are no \textbf{overstable} modes with $Im(\omega)>0$ and $Re(\omega)\neq0$. Therefore, for unstable modes, we set $\omega=i\gamma$ in the dispersion relation \eqref{jmog25}, where $\gamma$ is positive real number. Finally, the dispersion relation takes the following form
\begin{equation}
\frac{k^2}{k_J^2}\left(1+\frac{\alpha\beta}{\frac{k^2}{k_J^2}+\beta}\right)^{-1}=1-\sqrt{\pi} x e^{x^2}\left(1- \mathrm{erf}[x]\right)
\end{equation}

where $x=\frac{\gamma}{\sqrt{2}k\sigma}$. We have plotted this dispersion relation in Fig. \ref{de} for various values of $\beta$. In order to explain the rotation profile of spiral galaxies, $\alpha$ should satisfy the bound $\alpha=8.89\pm0.34$ \cite{rahvar}. In Fig. \ref{de} we have used $\alpha=8.89+0.34$. The dispersion relation of fluids, Eq. \eqref{jmog20}, is also plotted in Fig. \ref{de}. In fact we have plotted $\frac{\omega^2}{4\pi G \rho}$ with respect to $\frac{k^2}{\tilde{k}_J^2}$ for unstable modes ($\omega^2<0$). As long as $\frac{k^2}{\tilde{k}_J^2}<1$, $\omega^2$ is negative and the given mode is unstable. For $\frac{k^2}{\tilde{k}_J^2}=1$, $\omega^2$ vanishes, and for $\frac{k^2}{\tilde{k}_J^2}>1$, $\omega^2$ is positive. This means both the stellar and the fluid system are unstable if $k<\tilde{k}_J$. The corresponding limits of instability in Newtonian limit of GR also are plotted for both the stellar and the fluid system (curves with $(\alpha,\beta)=(0,0)$). As illustrated in Fig. \ref{de}, the limit of instability in MOG and Newtonian limit of GR are different. More specifically, in the case of fluid system (dashed curves), this limit in MOG is lower than the Newtonian case. However, for stellar systems the instability limit can be higher or lower than the Newtonian case.

\begin{figure}[!t]
   \includegraphics[width=0.4\textwidth,natwidth=250,natheight=250]{de.eps}
       \caption{Dashed curves show the dispersion relation for fluid systems. From left to right dashed curves correspond to $(\alpha,\beta)=(0,0)$, $(9.23,0.05)$, $(9.23,0.5)$ and $(9.23,1.3)$. The solid curves show the dispersion relation for stellar systems and from left to right they correspond to $(\alpha,\beta)=(9.23,1.3)$, $(0,0)$, $(9.23,0.5)$ and $(9.23,0.05)$.}
       \label{de}
\end{figure}

\begin{table*}[!t]
\begin{center}
\caption{Characteristics of the different phases of the ISM. For each phase, we have calculated the Jeans mass in Newtonian limit of GR and MOG. Since in MOG the difference between $M_J$ and $\tilde{M}_J$ is quite small for most phases, we have shown the fractional difference $(\Delta\tilde{M}_J/M_J)_{MOG}$ instead of $\tilde{M}_J$. For all cases, we have used the average value $\mu=2$ for the mean molecular weight.}
\begin{tabular}{c c c c c c c c c}

\hline\hline
Phase &~~~~ $n(cm^{-3})$ &~~~~ $T$(K)  & ~~~~$M_J$($M_{\odot}$)  &~~~~ $\lambda_J$(pc)& ~~~~$(\Delta\tilde{M}_J/M_J)_{MOG}$ & ~~~~ \\ \hline

Bok globules  &~~~~ $10^{4}$ & ~~~~$10$   & ~~~~$3.9$&~~~~$0.24$&~~~~$3.64\times 10^{-11}$ &~~~~ \\

Giant molecular clouds  &~~~~ $100$ & ~~~~$10$   & ~~~~$39$&~~~~$2.47$&~~~~$3.64\times 10^{-9}$ &~~~~ \\

Cold neutral medium  &~~~~ $30$ & ~~~~$80$   & ~~~~$1.61\times 10^3$&~~~~$12.75$&~~~~$9.32\times 10^{-8}$&~~~~  \\

Warm neutral medium  &~~~~ $0.6$ & ~~~~$8\times10^3$  & ~~~~$1.14\times 10^7$&~~~~$9.02\times 10^2$&~~~~$4.66\times 10^{-4}$ &~~~~ \\

Warm ionized medium &~~~~ $0.1$ & ~~~~$8\times 10^3$   & ~~~~$2.8\times 10^{7}$&~~~~$2.21\times10^3$&~~~~$2.89\times 10^{-3}$ &~~~~ \\

HII regions &~~~~ $0.1$ & ~~~~$10^4$   & ~~~~$3.9\times 10^{7}$&~~~~$2.5\times 10^3$&~~~~$3.4\times 10^{-3}$ &~~~~ \\

Hot intercloud &~~~~ $0.004$ & ~~~~$10^6$   & ~~~~$1.95\times10^{11}$&~~~~$1.24\times 10^5$&~~~~$0.78$ &~~~~ \\

 \hline\hline
\label{table1}

\end{tabular}
\end{center}
\end{table*}

\section{The Jeans mass limit in interstellar and intergalactic medium} \label{MT}

In this paper the Jeans analysis and its extension was investigated in the context of MOG. In order to compare the new Jeans mass limit with the standard one, let us first briefly review some components of the interstellar medium (ISM).

\begin{enumerate}
  \item  One of the most popular elements of ISM is the \emph{Giant Molecular Clouds} (GMCs). GMCs are enormous complexes of gas, dust and substructures with total masses in the range $10^4-10^6 M_{\odot}$, average densities of $10^2-10^3 cm^{-3}$ and the temperature T$\sim$ 10 K. Their size is of the order of 10pc.
    Inside of GMC there are some kind of denser substructures such as:
   \emph{Clumps}, \emph{Dense cores} and \emph{Hot cores}.
  \item   \emph{Bok globules} are one of the most known star forming regions. They are small interstellar clouds of very cold gas and dust that located outside of large molecular complex. They have almost spherical geometry and  are optically thick for visible light. They are characterized by very low temperature, relatively large number density ($n \ge 10^{4} cm^{-3}$), low masses (1 to 1000 $M_{\odot}$ ) and small size roughly around 1 pc. Bok globules are relatively isolated, and normally contain a dense core. Therefor, they may be the precursors to protostars. Recent observations of Bok globules revealed the inflow and outflow structure which are common process in the development of a protostar. These observations support the theory of star formation inside an isolated Bok globule. Bok globules are particularly interesting to astronomers since they are not too far, four times closer than the closest giant complexes, which allowing us more detailed observations. So They are the best candidate for testing star-formation scenarios.

\item \emph{HII regions} are one of the most easily seen and spectacular object in the galaxy. Cosmic-ray protons, intense UV emission  from O and early B-type stars which have formed inside the molecular cloud and shock waves from supernovae regions, are the sources for ionizing hydrogen in the HII regions. HII regions can be quite large. O stars can typically ionized a region roughly around hundred of parsecs while B stars can only ionized a region on the order of a few parsecs. Overall masses of HII regions are of the order of $10^2 -10^4 M_{\odot}$. Typical number density in the bright parts of HII regions is of the order of $0.1-10^4 cm^{-3}$. The temperature of this region is of the order of $10^4 K$.

\item \emph{Warm ionized mediums:}  Extended low-density photo-ionized regions which often referred to as the warm ionized medium, contain much more total mass than the more visually conspicuous high-density HII regions. This warm ionized medium has low density $0.1~cm^{-3} $ and  relatively high temperature $T \sim 8000 K$.

\item \emph{Cold neutral mediums} consist of neutral hydrogen (HI) and molecules at temperature $T\sim 10-100 K$ and relatively low number density $ 30 ~ cm^{-3}$.

\item \emph{Warm neutral mediums} consist of natural hydrogen HI and molecules at temperatures $T\sim 10^{3}-10^4 K$ with very low number density $0.6~cm^{-3}$.

\item \emph{Hot interclouds} consist of ionized gas (HII) at relatively high temperature $T\sim 10^{5}-10^6 K$. This regions are the lowest density phase in the ISM, $n\sim 0.004~cm^{-3}$, with the typical size roughly around $\sim 20~pc$.

\end{enumerate}

\begin{table*}[!t]
\begin{center}
\caption{Properties of common quasar absorption line systems.}
\begin{tabular}{c c c c c c c c c}
\hline\hline
 Absorber class &~~ $n(m^{-3})$ & $T$($10^3$K)  &$R$(kpc)& ~~$M_J$($10^9 M_{\odot}$)  & $\lambda_J$(kpc)& $(\Delta\tilde{M}_J/M_J)_{MOG}$  \\ \hline

Damped Ly$\alpha$ Absorbers& $ 10^4-10^7$ &~~~  $ 0.1-10$  &  $10-20$ &   $3.9\times10^{-6}-0.123$ &~~ $0.025-7.81$ &~~$3.64\times 10^{-6}-0.03$  \\
Super LLS & $10^4$ & $10$  & $-$ & $0.123$&$7.81$&$0.03$  \\
Lyman Limit Systems & $10^3-10^4$ & $30$  & $-$ & $0.641-2.03$&  $130-430$& $0.09-0.44$  \\
Ly$\alpha$ forest & $0.01-10^3$ & $5-50$  & $15-100$ & $4.36-43.6$&$55.2-5520$&$0.54-0.97$  \\
 \hline\hline
\label{table2}
\end{tabular}
\end{center}
\end{table*}

For the ISM the velocity dispersion $\sigma$ of the particles due to the temperature is expressed in terms of the temperature $T$ and the mean molecular mass $\mu m_p$ as
\begin{equation}
\sigma^2=\frac{k_B T}{\mu m_p}
\label{sigma1}
\end{equation}
where $\mu$ is the mean molecular weight, $k_B$ is the Boltzmann constant and $m_p$ is the proton mass. Also the density $\rho_0$ can be written as
\begin{equation}
\rho_0=\mu m_p n
\label{rho1}
\end{equation}
where $n$ is the number density of the particles. Using Eq. \eqref{sigma1} and \eqref{rho1}, the standard Jeans mass takes the following form
\begin{equation}
M_J\sim\frac{155.98}{\mu^2}M_{\odot} \left(\frac{T}{10K}\right)^{3/2}\left(\frac{n}{10^4 cm^{-3}}\right)^{-1/2}
\label{jmass2}
\end{equation}
and the new Jeans mass is given by \eqref{jmass}. In this case, the dimensionless parameter $\beta$ is
\begin{equation}
\beta=\frac{k_B\mu_0^2}{4\pi G m_p^2\mu^2}\frac{T}{n}\sim \frac{10^{-7}}{\mu^2}\left(\frac{T}{10K}\right)\left(\frac{n}{10^4 cm^{-3}}\right)^{-1}
\end{equation}
where we have used $\mu_0=0.042$ $kpc^{-1}$ \cite{rahvar}. In Table \ref{table1}, we have listed some physical properties of different ISM phases. Also, the Jeans mass $M_J$, the Jeans wavelength $\lambda_J$ and the fractional difference $(\Delta \tilde{M}_J/M_J)_{MOG}$ are calculated. It should be noted that the fractional difference is defined as
\begin{equation}
\frac{M_J-\tilde{M}_J}{M_J}=1-\sqrt{8}\left(1-\beta+\sqrt{(1+\beta)^2+4\alpha\beta}\right)^{-3/2}
\end{equation}
It is clear from Table \ref{table1} that, except for the Hot interclouds, the difference between $M_J$ and $\tilde{M}_J$ is small. In the case of metric $f(R)$ gravity studied in \cite{capo}, the fractional difference $(\Delta \tilde{M}_J/M_J)_{f(R)}$ is the same for all phases and equals to $0.296$.

However, on another scale, the low densities and high temperatures of some
regions of Intergalactic Medium (IGM) could yield to large Jeans masses comparable to the masses
of galaxies. We shall see that in the IGM the difference between $M_{J}$ and $\tilde{M}_J$ can be quite large.

Again, in order to compare $M_J$ and $\tilde{M}_J$ in the intergalactic space, let us briefly review the relevant physical properties of the IGM. The IGM is the baryonic material filling the space between galaxies. This is the main baryonic component of the universe and it is believed that the IGM is the material from which galaxies form. Also, it is known that quasars spectrum contains valuable information about the physical properties of the intergalactic space \cite{lynds1971}. In fact, the light from a distant quasar travels large distances before reaching us. Any intergalactic cloud which lies along the line-of-sight from the observer to the quasar imprints its signature on the spectrum of the quasar in the form of absorption lines. The clouds that generate individual absorption features are specified with the name of \emph{Lyman alpha systems} (Ly$\alpha$). Ly$\alpha$ systems are grouped into several categories \cite{igm}:

\begin{enumerate}
  \item \emph{Ly$\alpha$ forest}: The redshift of these systems lies between $1.5<z<4$. Their
   typical metallicity is $Z \sim 10^{-2} Z_{\odot}$, where $Z_{\odot}$ is the metalicity of the Sun. Also, for these systems $ 10^{12}<N_{HI} < 10^{17} cm^{-2}$. Where $N_{HI}$ is called the column density and defined as the number of
   hydrogen atoms per unit area (i.e. the number density integrated along the line-of-sight through the absorbing system).

  \item \emph{Lyman Limit systems (LLSs) }: Systems with column densities $10^{17} < N_{HI} < 10^{19} cm^{-2}$ and reshifts  $0.32 < z < 4.11$ are called Lyman-limit systems.  They exhibit an obvious discontinuity at the Lyman limit \cite{igm}.

\item \emph{Super Lyman Limit Systems (sLLSs)} are systems with $10^{19} < N_{HI} < 2 \times 10^{20} cm^{-2}$. These systems are more convenient for statistical studies since their column densities can be easily determined.
\item \emph{Damped Ly$\alpha$ Absorbers} are
 systems with highest column densities $N_{HI}> 10^{20} cm^{-2}$. Their redshift is in the range $0.1<z<4.7$. Their typical metallicities range between 0.02 and 0.1 $Z_{\odot}$.

\end{enumerate}

In Table \ref{table2}, we have listed the physical properties of Ly$\alpha$ systems \cite{igm}. Also, the Jeans mass and the corresponding Jeans wavelength are calculated. It is clear from this table that the new Jeans mass can be dramatically smaller than the standard Jeans mass. For example, in LLS systems the standard Jeans mass can be 44\% larger than the new Jeans mass.  In the Ly$\alpha$ forest systems this ratio can be even larger and reaches to 97\%. On the other hand, the difference between $M_J$ and $\tilde{M}_J$ in Super LLS and Damped Ly$\alpha$ absorbers is not too large.

\section{discussion and conclusions}\label{DC}

In this paper we have studied the Jeans instability in the context of MOG. By considering the Newtonian limit of MOG, we find the corresponding Poisson and Boltzmann equations. Finally linearizing the governing equations of the fluid and stellar systems, we derived the gravitational instability criterion. In fact, we found new Jeans mass limit $\tilde{M}_J$ \eqref{jmass}. The main theoretical result of this work is that the new Jeans mass is slightly below the standard one and contains two parameters $\alpha$ and $\mu_0$. The magnitude of these parameters has been fixed from the rotation curve data \cite{rahvar}. In section \ref{MT}, we have calculated the new Jeans mass for different components of interstellar and intergalactic medium.

The Jeans instability in ISM is directly related to star formation. Therefore, comparing $M_J$ and $\tilde{M}_J$ may help us to compare Newtonian dynamics and MOG in this astrophysical level (star formation).

The gravitational instability is one of dominant physical process in star formation theory. It is known that stars form within Molecular Clouds. However, the detailed analysis of star formation process is a difficult theoretical problem. It is well understood now that the star formation is not simply the result of giant clouds breaking apart into tiny, dense substructures. The onset of collapse is rather a highly localized occurrence within large complexes, and dictated by the initial condition. Also other processes including fragmentation, reduction of angular momentum and magnetic flux intrinsic to the ionized gases of the ISM are important in star formation processes.

A molecular cloud will remain in hydrostatic equilibrium, until as a consequence of their low temperatures and high densities it exceeds its Jeans mass and starts to collapse. Perhaps a better way to understand the difference between molecular clouds, clumps, hot cores
and Bok globules is by determining their Jeans mass and length scales. In this case, using Eq \eqref{jmass2} one can easily verify that the Jeans length is given by

\begin{equation}
\lambda_J=\frac{1.06pc}{\mu^2}  \left(\frac{T}{10K}\right)^{1/2}\left(\frac{n}{10^4 cm^{-3}}\right)^{-1/2}
\label{la}
\end{equation}
which gives us an estimate for the minimum initial separation for self-gravitating fragments. Thus any dense molecular core containing more than a few tens of solar masses of gas is unstable, and will collapse in a free-fall time scale.

As the cloud becomes unstable it will start to collapse and consequently its density will increase, but the atomic and molecular processes cooling it effectively. So as long as the cloud is optically thin, the temperature within the cloud remains roughly constant. As a result, the Jeans mass steadily decreases as the collapse proceeds, and in the early stage of star formation process, collapsing cloud fragments into lower and lower mass pieces, each collapsing on its own free-fall time scale. Therefore, initial collapsing cloud is on its way to becoming a star cluster.

In the ISM the fractional difference can be written as
\begin{equation}
\begin{split}
\frac{\Delta\tilde{M}_J}{M_J}&\sim\frac{3}{2}\alpha\beta= \frac{3\alpha k_B\mu_0^2}{8\pi G m_p^2\mu^2}\frac{T}{n} \\ & \sim \frac{1.5\times10^{-7}\alpha}{\mu^2}\left(\frac{T}{10K}\right)\left(\frac{n}{10^4 cm^{-3}}\right)^{-1}
\end{split}
\label{nn}
\end{equation}
In the case $\alpha=8.89$, small value of $\beta$ causes a small deviation between $M_J$ and $\tilde{M}_J$. It is also clear from Eq. \eqref{nn} that for systems with high temperature and low number density $\beta$ can be large enough to provide a significant difference between $M_J$ and $\tilde{M}_J$. This is the case for Hot interclouds, see Table \ref{table1}.

As it is shown in Table \ref{table1}, except for the hot interclouds, the difference between $M_J$ and $\tilde{M}_J$ in the ISM is quite small. Taking into account that hot interclouds are not effective star forming sites, one can infer that there is not significant difference between star formation scenarios in MOG and GR. This result can be simply justified by comparing the parameter $\mu_0^{-1}$ with the Jeans length of the given medium. Using Eq. \eqref{la} and the characteristic values for the number density and temperature of the different phases of the ISM, one can show that $\mu_0\lambda_J\ll1$, or equivalently $\beta\ll1$. This means that on the ISM scale, the modification induced by MOG to the gravitational law, is very small. In other words, as expected, for seeing a substantial difference between MOG and GR one should compare them in larger self-gravitating systems.

However, on another scale in IGM, where the galaxies form, the difference between $M_J$ and $\tilde{M}_J$ can be significantly large. This means that although in the case of star formation there is not an important difference between Newtonian physics and MOG, but there may be substantial differences between these theories in the galaxy formation scenarios. Furthermore, one may expect that the cosmological structure formation in the context of MOG may be different from the standard cosmological model \cite{me}. In order to make this possibility more clear, let us find the Jeans mass in the epoch of equality of matter and radiation energy densities (at redshift $z_{eq}$). At this epoch the adiabatic sound speed is given by
\begin{equation}
c_s=\frac{2}{\sqrt{21}}c
\end{equation}
where $c$ is the velocity of light. Therefore the Jeans wavelength and mass are given by
\begin{equation}
\begin{split}
&\lambda_J^{eq}=2.4 \left(\frac{c h}{H_0}\right)(\Omega_{m0}h^2)^{-1/2}(1+z_{eq})^{-3/2}\\&
M_J^{eq}= 0.7 \left(\frac{c^3 h}{H_0 G}\right)(\Omega_{m0}h^2)^{-1/2}(1+z_{eq})^{-3/2}
\end{split}
\label{nn1}
\end{equation}
where $\Omega_{m0}=\rho_{m0}/\rho_c$ is the current value of the dimensionless matter density parameter, $\rho_{m0}$ is the energy density of matter at present and $\rho_c=3H_0^2/8\pi G$ is the critical density. Also $h$ is the Hubble dimensionless parameter ($H_0=100 h~km~s^{-1}~Mpc^{-1}$). Eqs \eqref{nn1} can be easily verified using the definitions of $\lambda_J$ and $M_J$ and keeping in mind that at the equality epoch the energy density of the matter is $\rho_m(z_{eq})=\rho_{m0}(1+z_{eq})^3$. If we use the recent results of \emph{Planck} for cosmological parameters \cite{planck}, i.e. $\Omega_{m0}h^2= 0.143, h=0.6711$ and $z_{eq}=3402$, then we get
\begin{equation}
\lambda_J^{eq}\sim 89.48~kpc~~~~~~~~~~M_J^{eq}\sim 5.87\times 10^{17} M_{\odot}
\end{equation}
Thus at this epoch the Jeans mass is of the same order of the mass of a supercluster. In this case $\beta=0.36$ (for $\mu_0=0.042 kpc^{-1}$ and $\alpha=8.89$) and the corresponding Jeans length and mass in MOG are
\begin{equation}
\tilde{\lambda}_J^{eq}\sim 0.67~\lambda_J^{eq}~~~~~~~~~~\tilde{M}_J^{eq}\sim 0.30 ~M_J^{eq}
\end{equation}
Therefore at this epoch the new Jeans mass is 70 \% smaller than the standard one. Therefore one may expect a significant difference between structure formation process in the context of MOG and the standard $\Lambda$CDM model \cite{me}. It should be stressed that we made an order of magnitude estimation for the new Jeans mass at the equality epoch. It is obvious that for studying the structure formation in the context of MOG, one should start with the relevant equations in the expanding universe. 
\section{Acknowledgment}
This work is supported by Ferdowsi University of Mashhad under
the grant 2/28687 (28/08/1392). We would like to thank Hossein Haghi and Mariaflecia~De Laurentis for helpful discussions and comments.


\begin{thebibliography}{99}
\bibitem{berton}
G.~Bertone, \emph{Particle Dark matter, Observations, Models and Searches} (Cambridge
University Press, 2010).





\bibitem{lux}
  D.~S.~Akerib {\it et al.}  [LUX Collaboration],
  Phys.\ Rev.\ Lett.\  {\bf 112}, 091303 (2014).

\bibitem{LUX2}
  E.~Aprile {\it et al.}  [XENON100 Collaboration],
  Phys.\ Rev.\ Lett.\  {\bf 109}, 181301.

\bibitem{mofi}
J.~W.~Moffat, JCAP {\bf 0603}, 004 (2006).

\bibitem{bro1}
J.~R.~Brownstein and J.~W.~Moffat, Mon.\ Not.\ Roy.\ Astron.\ Soc. {\bf 367}, 527 (2006).

\bibitem{rahvar}
  J.~W.~Moffat and S.~Rahvar,
  Mon.\ Not.\ Roy.\ Astron.\ Soc.\  {\bf 436}, 1439 (2013).

\bibitem{bro2}
J.~R.~Brownstein and J.~W.~Moffat, Astrophys.\ J. {\bf 636}, 721 (2006).


\bibitem{mofi3}
J.~R.~Brownstein and J.~W.~Moffat,  Mon.\ Not.\ Roy.\ Astron.\ Soc. {\bf 382}, 29 (2007).

\bibitem{capo}
S.~Capozziello, M.~De Laurentis, I.~De Martino, M.~Formisano and S.D.~Odintsov, Phys.\ Rev. \ D {\bf 85}, 044022 (2012).

\bibitem{capo100}
S. Capozziello, M. De Laurentis, S. D. Odintsov and A. Stabile, Phys. Rev. D 83, 064004 (2011). 
  
\bibitem{dolgov}  
E.~V.~Arbuzova, A.~D.~Dolgov and L.~Reverberi,
  arXiv:1406.7104 [gr-qc].
  \bibitem{clifton}
  T.~Clifton, P.~G.~Ferreira, A.~Padilla and C.~Skordis,
  Phys.\ Rept.\  {\bf 513}, 1 (2012)
\bibitem{dombriz} 
G.~J.~Olmo,
  Phys.\ Rev.\ D {\bf 75}, 023511 (2007); J.~A.~R.~Cembranos, A.~de la Cruz-Dombriz and B.~Montes Nunez,
  JCAP {\bf 1204}, 021 (2012); 
  F.~Briscese and E.~Elizalde,
  Phys.\ Rev.\ D {\bf 77}, 044009 (2008); A.~de la Cruz-Dombriz, A.~Dobado and A.~L.~Maroto,
  Phys.\ Rev.\ D {\bf 80}, 124011 (2009).
\bibitem{haghi}
R.~H.~Sanders,
  Mon.\ Not.\ Roy.\ Astron.\ Soc.\  {\bf 296}, 1009 (1998); 
  M.~Malekjani, S.~Rahvar and H.~Haghi,
  Astrophys.\ J.\  {\bf 694}, 1220 (2009).
  
\bibitem{PRDme}
Mahmood Roshan, Phys.\ Rev. \ D {\bf 87}, 044005 (2013).

\bibitem{cqg2009}
J.~W.~Moffat and V.~T.~Toth, Class. Quant. Grav. {\bf 26}, 085002 (2009).

\bibitem{willbook}
Clifford~M.~Will, \textit{Theory and Experiment in Gravitational
Physics} (Cambridge University Press, Cambridge,
England, 1993).

\bibitem{papa}
A.~Papapetrou, Proc.\ Roy.\ Soc.\ London A {\bf 209}, 248 (1951).


\bibitem{will}
  B.~Bertotti, L.~Iess and P.~Tortora,
  Nature {\bf 425}, 374 (2003).

\bibitem{binney}
J.~Binney and S.~Tremaine, \textit{Galactic Dynamics} (Princeton University Press, Princeton, 1994).

\bibitem{milgrom}
M.~Milgrom,  Astrophys.\ J. {\bf 270}, 365 (1983).

\bibitem{capo3}
S.~Capozziello, V.~Faraoni, \textit{Beyond Einstein Gravity} (Springer, 2011).



  \bibitem{lynds1971}
  R.~Lynds, 1971, Astrophys. J. 164, L73+.


\bibitem{igm}
 A.~A.~Meiksin,
  Rev.\ Mod.\ Phys.\  {\bf 81}, 1405 (2009).


  \bibitem{me}
  Mahmood Roshan and Shahram Abbassi, work in progress
    \bibitem{planck}
      P.~A.~R.~Ade {\it et al.}  [Planck Collaboration],
  ``\emph{Planck 2013 results. XVI. Cosmological parameters},''
  arXiv:1303.5076.
\end{thebibliography}
\end{document}